\documentclass[aps,amsfonts,amsmath,amssymb]{revtex4-1}

\pdfoutput=1

\usepackage{hyperref}
\usepackage[utf8x]{inputenc}	
\usepackage[T1]{fontenc}
\usepackage{mathrsfs}
\usepackage{color}
\usepackage{fancyhdr}
\usepackage{verbatim}
\usepackage{graphicx}
\usepackage[]{natbib}

\def\w{\omega}

\def\Re{\mathcal{R}\mathrm{e}}
\def\R{\mathbb{R}}

\def\l{\lambda}

\begin{document}

\title{Broadband acoustic isolator in the audible range : nonreciprocal acoustic propagation using programmable boundary conditions}
\author{Sami Karkar\footnote{sami.karkar@ec-lyon.fr}}
\affiliation{LTDS, \'Ecole Centrale de Lyon -- CNRS UMR5513, 36 avenue Guy de Collongue, 69134 Ecully Cedex (France)}
\author{Etienne Rivet}
\affiliation{Acoustic Group, Laboratory of Signal Processing 2, IEL-STI, \'Ecole polytechnique fédérale de Lausanne, 10 route cantonale, 1015 Lausanne (Switzerland)}
\author{Manuel Collet}
\affiliation{LTDS, \'Ecole Centrale de Lyon -- CNRS UMR5513, 36 avenue Guy de Collongue, 69134 Ecully Cedex (France)}


\begin{abstract}
We propose a novel concept of nonreciprocal devices in acoustics, illustrated by the design of an acoustic diode, or isolator. A boundary control strategy was previously shown to provide direction-dependent propagation properties in acoustic waveguides. In this paper, the boundary control is reinterpreted as a source term for the inhomogeneous wave equation, in a purely 1D model. Nonreciprocity is then obtained using a distributed source that replaces the non-standard boundary condition 
where the normal velocity at the boundary is a function of both pressure and its tangential derivative. Numerical simulations are carried out to validate the theoretical model, and the scattering matrix of the device is retrieved to investigate the nonreciprocal nature of the system. Results show that the proposed device can lead to an efficient, ultra-broadband, sub-wavelength, acoustic isolator. Finally, the effects of actual, finite-sized transducers on the performance of the acoustic isolator are discussed.
\end{abstract}

\maketitle

\section{Introduction}

In a classical medium, acoustic waves behave identically in one direction or the opposite: the propagation of such waves has the property of reciprocity, owing to the symmetrical nature of physical laws under time reversal. Reciprocity is also observed by many other wave propagation phenomena, such as electromagnetism, mechanics and acoustics. Though uncommon in nature, direction-dependent propagation is sometimes a desirable feature: it provides isolation capabilities, which is of particular interest in all fields of physics. In electromagnetism, it has been shown that reciprocity can be broken in several ways: using biasing techniques like the Faraday effect \cite{aplet:1964}, nonlinear techniques \cite{krause:2008,poulton:2012,tocci:1995,gallo:2001}, or time-dependent modulation of properties \cite{lira:2012,yu:2009}.

The acoustic isolator, often called acoustic diode, is a device that let acoustic waves propagate in one direction, but blocks the transmission of acoustic waves in the reverse direction, similarly to an optical isolator \cite{jalas:2013}. Although the term ``diode'' is often used to name this type of system, the isolator differs from the usual electrical diode, in the electrical-acoustical direct analogy where the electrical current $i$ is associated to the acoustical velocity $v$: that would lead to a \emph{rectifying} diode, which would let the positive half-period of an harmonic acoustic wave pass, but stop the negative other half-period, generating DC and second harmonic components (note that such a diode has been patented long ago \cite{brevet_diode}).

Different concepts of acoustic isolators have been proposed, relying on strongly nonlinear media with higher harmonic conversion \cite{liang:2009,liang:2010,boechler:2011,popa:2014}, static bias such as circulating fluid \cite{Fleury:2014} (the acoustic equivalent of magnetic biasing in EM), or time-varying properties \cite{fleury:2015}. Those are actually the three known means to break reciprocity: nonlinearities, biasing, or time-varying properties. Some concepts relying solely on geometric features have been proposed, but taking into account all the inputs and outputs of the system shows that such concepts actually do not break reciprocity \cite{Maznev:2013}.

Most nonlinearity-based devices are usually limited by the necessary high input levels for the up-frequency conversion to be efficient. Very poor transmission is obtained in the transmit direction otherwise. On the other hand, biased, resonant-type devices are inherently very narrow-band. The present paper aims at addressing these two points, using active elements and the concept of programmable boundary control and adaptive generalized impedance.

The outline of this contribution is as follows. First, from a general 3-dimensional model of guided waves in a duct, we formulate a one-dimensional model, where the direction-dependent boundary control of non-locally reacting walls comes down to a distributed source. Then, we study a finite-length one-dimensional (1D) acoustic isolator based on this model, and characterize it in terms of its scattering matrix. Next, based on finite-element simulations, we propose numerical experiments to validate our model. Finally, we discuss the practical realization of the boundary control with distributed in-wall, flush-mounted actuators and sensors.

\section{Theory\label{sec:theory}}

\subsection{Three-dimensional acoustic propagation in a waveguide}
We consider a cylindrical waveguide with a cross-section of arbitrary shape $\Omega\in\R^2$, and area $S$. The longitudinal coordinate along the waveguide is denoted $x$, transverse directions are $(y,z)$. The closed contour around the cross-section is noted $\partial\Omega$.

\begin{figure}
	\centering
	\includegraphics[width=8.6cm]{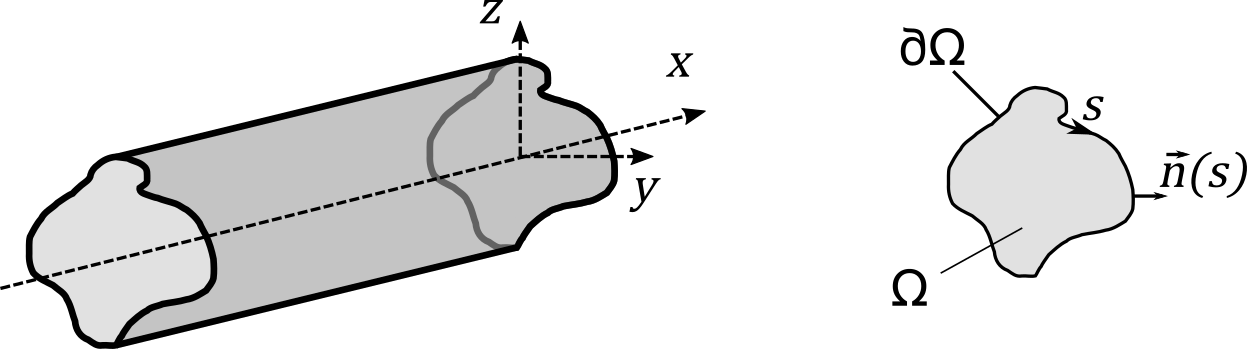}
	\caption{A cylindrical waveguide along coordinate $x$, with cross section of arbitrary shape $\Omega$. Left: overview of the waveguide. Right: detail of the cross-section and its contour $\partial\Omega$ parametrized by a curvilinear coordinate $s$. $\vec{n}$ is the local exterior normal at each point of the contour.}
\end{figure}

Inside such a waveguide, the linearized acoustic equations lead to the wave equation:
$\forall t\in \R, \; \forall x\in\R$, and $(y,z)\in\Omega$:
\begin{equation}
\frac{1}{c_0^2}\partial_{tt} p(x,y,z,t) - \nabla^2 p(x,y,z,t) = 0
\end{equation}
and the boundary conditions read:\\
$\forall t\in \R, \; \forall x\in\R$, and all $(y,z)\in\partial\Omega$,
\begin{equation}
\partial_n p(x,y,z,t) = -\rho_0 \partial_t \vec{v}(x,y,z,t).\vec{n} = -\rho_0 \partial_t v_n(x,y,z,t)
\end{equation}
where unit vector $\vec{n}$ is the local outgoing normal to the waveguide, $v_n$ is the outgoing normal velocity of the boundary of the waveguide (the ``wall''), $c_0=\sqrt{\chi/\rho_0}$ is the speed of sound in the host medium, with $\rho_0$ its mean density and $\chi$ its bulk modulus.

\subsection{Boundary control of wave propagation}

In case of a hard-walled waveguide, the normal velocity $v_n$ of the boundary is 0. In case of locally reacting walls with a given impedance, the normal velocity $v_n$ is proportional to the acoustic pressure $p$. This is typically the case of conventional absorbing materials, Helmholtz resonators, and other passive linings. In the case of an active lining \cite{karkar:2015a}, the normal velocity $v_n$ could be a more complex function of the acoustic pressure $p$ and its time or spatial derivatives. In the latter case, the moving wall could represent a generalized impedance, which is a non-local operator.

We now impose the following boundary condition on the walls of the waveguide, proposed by \citet{collet:2009}:
\begin{equation}
\rho_0 \partial_t v_n = -\partial_n p = \frac{1}{c_a} \partial_t p - \partial_x p
\label{bndcontrol}
\end{equation}
with $c_a$ an advection celerity, corresponding to a tunable parameter of the control. Although this partial derivative equation was meant as a boundary condition for the anomalous reflection of waves on an active surface, we use it here as a special boundary condition for a waveguide, where we expect waves to be under grazing incidence, in the low frequency regime.

\subsection{Reduction to 1D}
To reduce this model to one dimension, we average the equations by integrating over the cross-section:\\
$\forall t\in \R,  \quad x\in\R$,
\begin{equation}
\frac{1}{S} \iint_\Omega \big[ \frac{1}{c_0^2} \partial_{tt} p(x,y,z,t) - (\partial_{xx} + \partial_{yy} + \partial_{zz}) p(x,y,z,t) \big] dS = 0.
\end{equation}

Denoting the mean acoustic pressure over the section $\tilde{p}(x,t)$=$\frac{1}{S} \iint_\Omega p(x,y,z,t)$, and integrating by part the second term (or, equivalently, using the divergence theorem, or Stokes' theorem), the equation now reads:\\
$\forall t\in \R,  \quad x\in\R$,
\begin{equation}
\frac{1}{c_0^2} \partial_{tt} \tilde{p}(x,t) - \partial_{xx} \tilde{p}(x,t) = \frac{1}{S}\oint_{\partial\Omega} \partial_n p(x,s,t) ds
\end{equation}
where $s$ is a curvilinear coordinate along $\partial\Omega$ and $\partial_n$ is the gradient along the normal direction directed outward.

Inserting the boundary control equation \eqref{bndcontrol} in the right-hand side, the equation now reads:
\begin{equation}
\frac{1}{c_0^2} \partial_{tt} \tilde{p}(x,t) - \partial_{xx} \tilde{p}(x,t) = -\frac{1}{S}\oint_{\partial\Omega} \frac{1}{c_a} \partial_t p(x,s,t) - \partial_x p(x,s,t) ds
\end{equation}

Denoting the mean acoustic pressure field along the boundary contour $p_b(x,t)$=$\frac{1}{L_p} \oint_{\partial\Omega} p(x,s)ds$, where $L_p$ is the perimeter length of the cross-section, we finally get:
\begin{equation}
\frac{1}{c_0^2} \partial_{tt} \tilde{p}(x,t) - \partial_{xx} \tilde{p}(x,t) = \partial_t Q_m(x,t)
\end{equation}
where the source term in the right-hand side is:
\begin{equation}
\partial_t Q_m = -\frac{L_p}{S} \big( \frac{1}{c_a} \partial_t p_b (x,t) - \partial_x p_b(x,t) \big).
\label{src_term}
\end{equation}

\subsection{1D acoustic isolator}
We now make assumption of monomodal acoustic propagation inside the waveguide (plane waves), an assumption that is very often used and completely validated in the low frequency range. Thus, $p_b$=$\tilde{p}$=$p$. We will therefore drop the tilda and subscript $b$ in what follows.

We then simply rewrite inhomogeneous 1D wave equation:\\
$\forall t\in \R, \quad \forall x\in[0,L],$
\begin{equation}
\frac{1}{c_0^2}\partial_{tt} p(x,t) - \partial_{xx}p(x,t) = \partial_t Q_m(x,t)
\label{eq:helmholtz}
\end{equation}
with the source term defined by:
\begin{equation}
\partial_t Q_m = -\frac{1}{d} \big( \frac{1}{c_a} \partial_t p (x,t) - \partial_x p(x,t) \big),
\label{src_term1D}
\end{equation}
where $L_p/S=1/d$ is now a second tuning parameter of the control, together with the advection celerity $c_a$.

\subsection{Dispersion relation}
Assuming a time-harmonic wave propagation with angular frequency $\w$ and wavenumber $k$, we now seek solutions of the form:
\begin{equation}
p(x,t) = p_0 e^{j(\w t-kx)}.
\label{eq:wave}
\end{equation}
Inserting this ansatz in the special source distribution equation \eqref{src_term} and in the propagation equation \eqref{eq:helmholtz}, comes:
\begin{equation}
(j\w/c_0)^2 p(x,t) - (-jk)^2 p(x,t) = -\frac{1}{d}\big((j\w/c_a) p(x,t) -(-jk) p(x,t)\big)
\end{equation}
which, being true for all $t\in\R$ and all $x\in[0,L]$, implies the following relationship:
\begin{equation}
(j\w/c_0)^2 - (jk)^2 = -\frac{1}{d}(j\w/c_a+jk).
\label{eq:dispersion}
\end{equation}

In the special case where $c_a=c_0=c$ (advection celerity in tune with the sound speed), we get:
\begin{equation}
j\w/c(j\w/c+1/d) = jk(jk-1/d)
\end{equation}
which has two obvious solutions: either $k^{(-)}=-\w/c$, or $k^{(+)}=\w/c-j/d$.

The first solution shows that waves with $\Re(k)<0$ have a negative group velocity, which is equal to the phase velocity, thus this type of wave is propagating in the (-) direction ("backward waves"). Given the purely real wavenumber, they pass through the isolator without attenuation.

Inserting the second solution in the ansatz \eqref{eq:wave}, we write:
\begin{equation}
p(x,t) = p_0 e^{j\w(t-x/c)}e^{-x/d}.
\end{equation}
This second solution shows that waves with $\Re(k)>0$ propagate in the (+) direction ("forward waves"), given their positive group velocity, but with an exponential attenuation. It is equivalent to an evanescent guided mode
, just like a waveguide higher mode under its cutoff frequency.

We can now physically interpret the meaning of $d$: it is the typical attenuation length. If it is short enough, a good acoustic isolator can be designed: sound waves can only pass in one direction through the isolator.


\subsection{Active power delivered by the source}
Given the source term $Q_m$, the power injected by the source of the isolator into the acoustic domain is obtained as follows:
\begin{equation}
\mathcal{P}(x,\w) = \frac{1}{2}\Re \big[Q_m(x,\w) p^*(x,\w)/\rho_0 \big]
\label{eq:power_product}
\end{equation}
where $\mathcal{P}$ here stands for the acoustic power density, or acoustic intensity per unit length.

Using the ansatz \eqref{eq:wave} and the definition \eqref{src_term} of the source term, comes:
\begin{equation}
\mathcal{P}(x,\w) = -\frac{1}{2d} \frac{|p(x,\w)|^2}{\rho_0 c} \Re \big[\frac{j\w/c + jk}{j\w/c}\big]
\label{eq:power}
\end{equation}

Clearly, for backward waves, where $k$=$-\w/c$, the power density of the source vanishes because of the numerator of the rightmost fraction. The source is inactive.

For forward waves, inserting the dispersion relation \eqref{eq:dispersion} into Eq. \eqref{eq:power}, the power density is:
\begin{equation}
\mathcal{P}(x,\w) = -\frac{1}{d} \frac{p_0^2}{\rho_0 c} e^{-2x/d}
\end{equation}
This value is always negative, for every $x$ and every $\w$, so the source is always absorbing power: the system is unconditionally stable for these waves.

However, due to internal reflections, when a forward incident wave enters the isolator, the pressure field inside the isolator actually consists of a superposition of forward and backward waves. The incident, forward wave cannot occur separately from the reflected, backward wave. Even though backward waves alone do not contribute to the source term, it does have an incidence on the power balance in case of a superposition of both forward and backward waves, as it contributes to the pressure term of the product in Eq. \eqref{eq:power_product}. In that case, the power balance reads:
\begin{equation}
\mathcal{P}(x,\w) = -\frac{1}{d} \frac{1}{\rho c} \Re \big[p^{(+)}(x,\w) . (p(x,\w))^*\big]
\label{eq:Psource}
\end{equation}
where the total acoustic pressure $p(x,\w)=p^{(+)}(x,\w)+p^{(-)}(x,\w)$ is a superposition of two waves with opposite directions of propagation. It is not obvious, in that case, to prove that the source is passive, so this point will be investigated in the next section, by means of numerical simulations.

\section{Results}

\begin{figure}
\centering
\includegraphics[width=8.6cm]{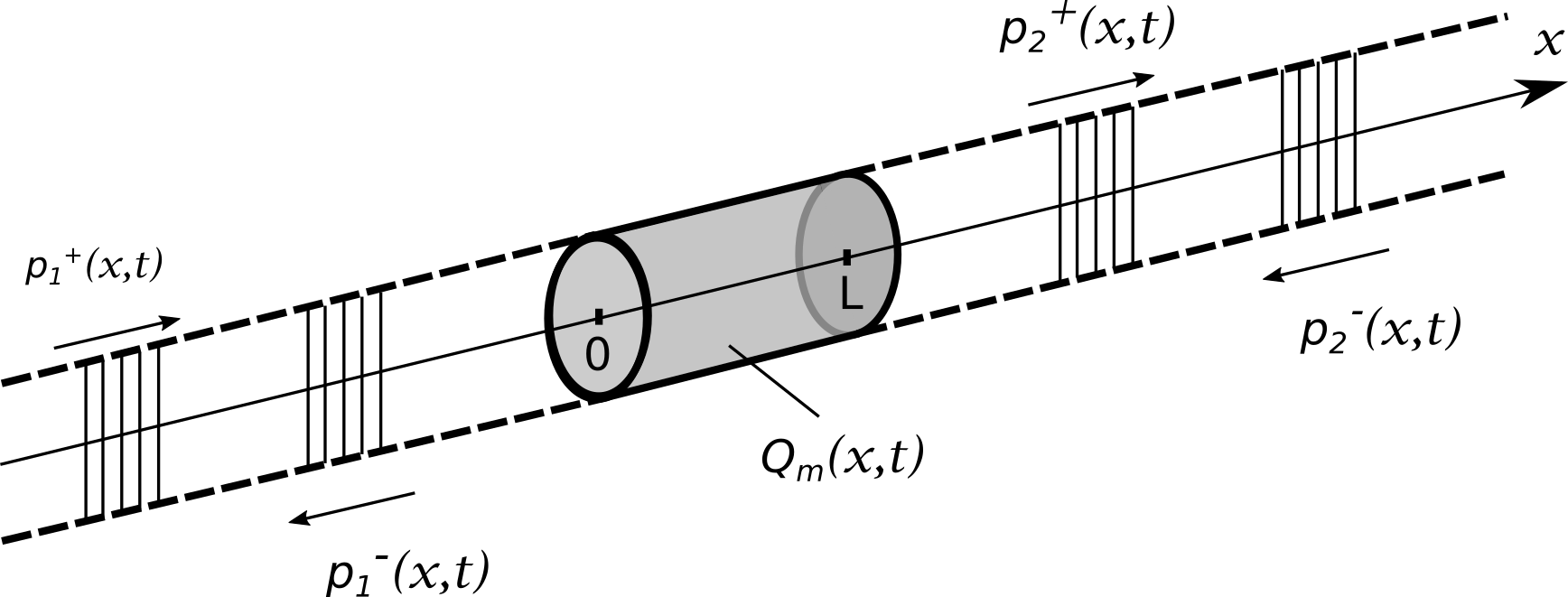}
\caption{A finite-sized acoustic isolator connected to semi-infinite waveguides at both ends. Ingoing and outgoing waves are shown on both sides.}
\label{fig:schema}
\end{figure}

We now simulate the theoretical model proposed in section \ref{sec:theory}. Using a finite element software in the frequency domain, and meshing the isolator uniformly with a step $h$=3.46mm (sufficient to resolves acoustic waves up to $\l$=$10h$, or $f$=10kHz), we solve for the wave equation on a portion of simple waveguide (a), connected to the isolator (b), terminated by a portion of simple waveguide (c), as depicted in figure \ref{fig:schema}. The medium is dry air at standard temperature and pressure: the density used for the medium is  $\rho_0$=1.18 kg/m$^3$ and speed of sound is $c$=346 m/s.

We apply a plane wave radiation boundary condition at the left-most and right-most ends of the system to realize perfect absorption, and we set $Q_m$ as defined by \eqref{src_term} on the isolator. A 1Pa RMS incident plane wave (94dB re. 20$\mu$Pa) is fed into the input waveguide (a), using a normal acceleration formulation. The total acoustic field is resolved in the whole system, for frequencies ranging from 10Hz to 10kHz.

\subsection{Pressure level profile}
\begin{figure}
	\centering
	\includegraphics[width=8.6cm]{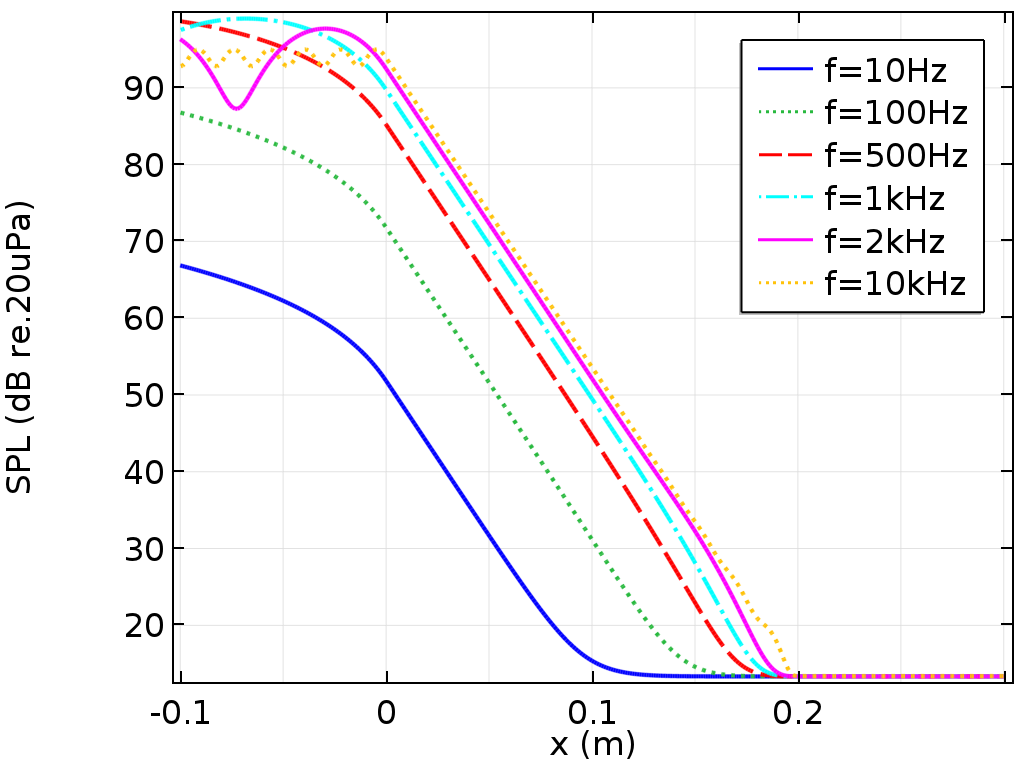}
	\caption{Blocked propagation for incident waves: sound pressure level (in dB re.20$\mu$Pa) along the system. Control parameter: $d$=2.2cm}
	\label{fig:1D-Lp}
\end{figure}
Figure \ref{fig:1D-Lp} shows the sound pressure level $L_p$ (in dB) along the system axis, for several frequency, when the control parameter is set to $d$=2.2cm. The linear decrease of $L_p$ with $x$ confirms the exponential decay of the incident waves inside the isolator. A reflected wave is created, resulting in interference patterns in the input waveguide, and frequency-dependent amplitude for the sound pressure level.

The equivalent figure for retrograde waves (not shown here), is perfectly flat: the waves propagate as in a passive waveguide.


\subsection{S-matrix}
\begin{figure*}
	\centering
	\includegraphics[width=.45\textwidth]{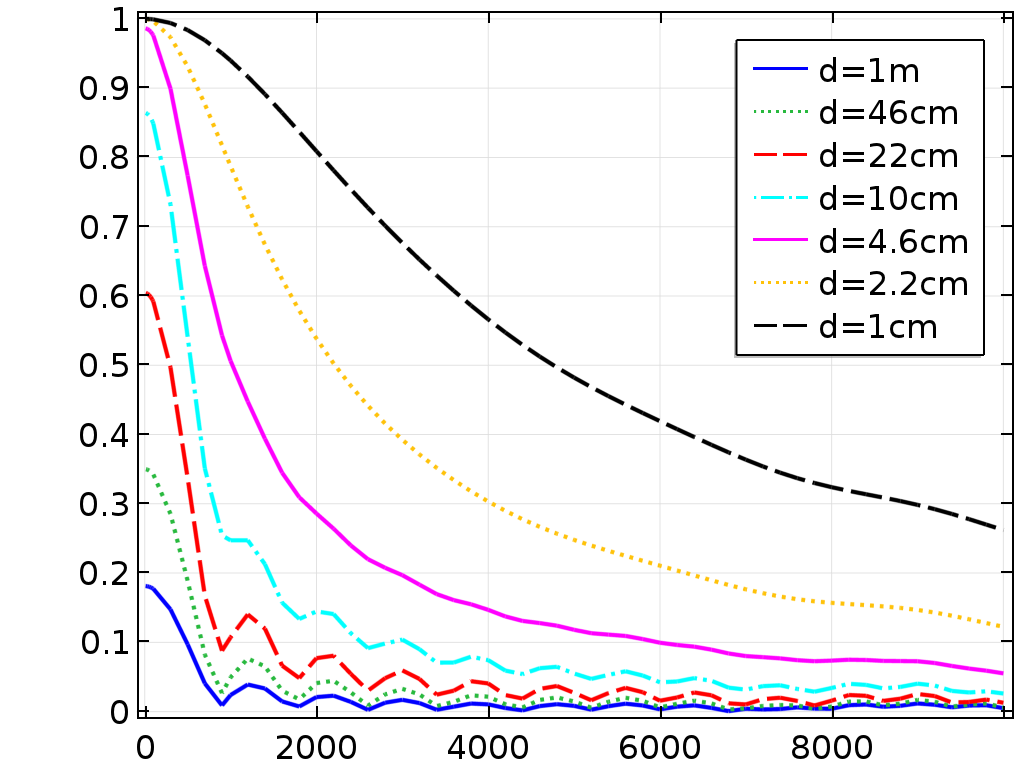}\includegraphics[width=.45\textwidth]{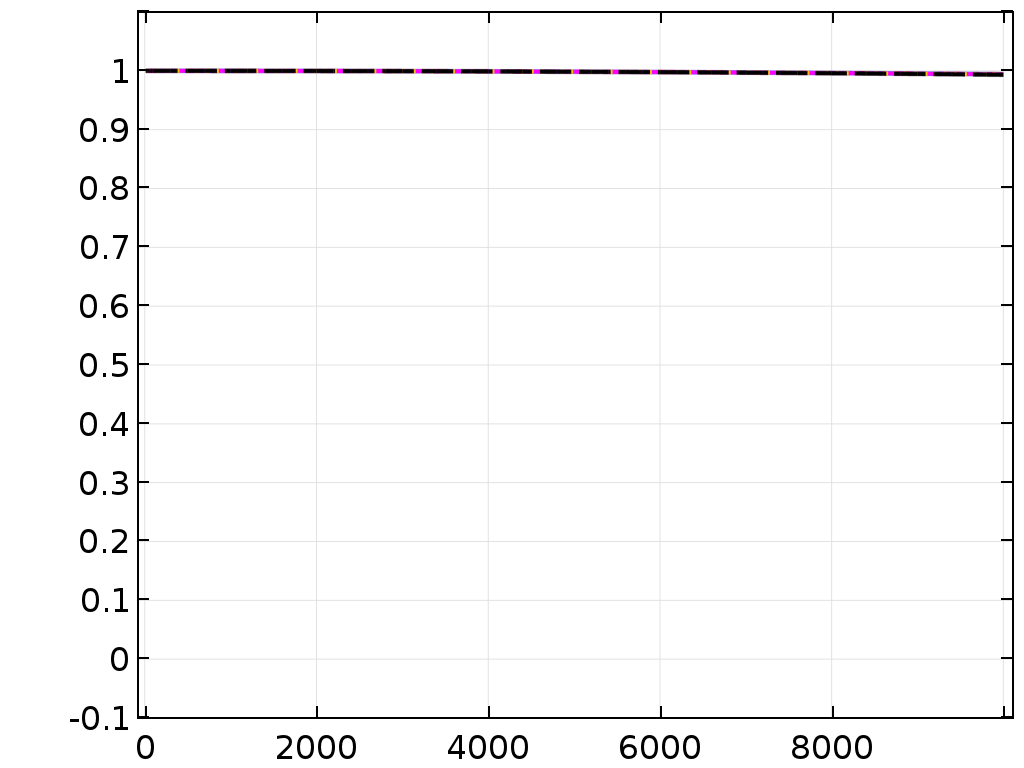}\\
	\includegraphics[width=.45\textwidth]{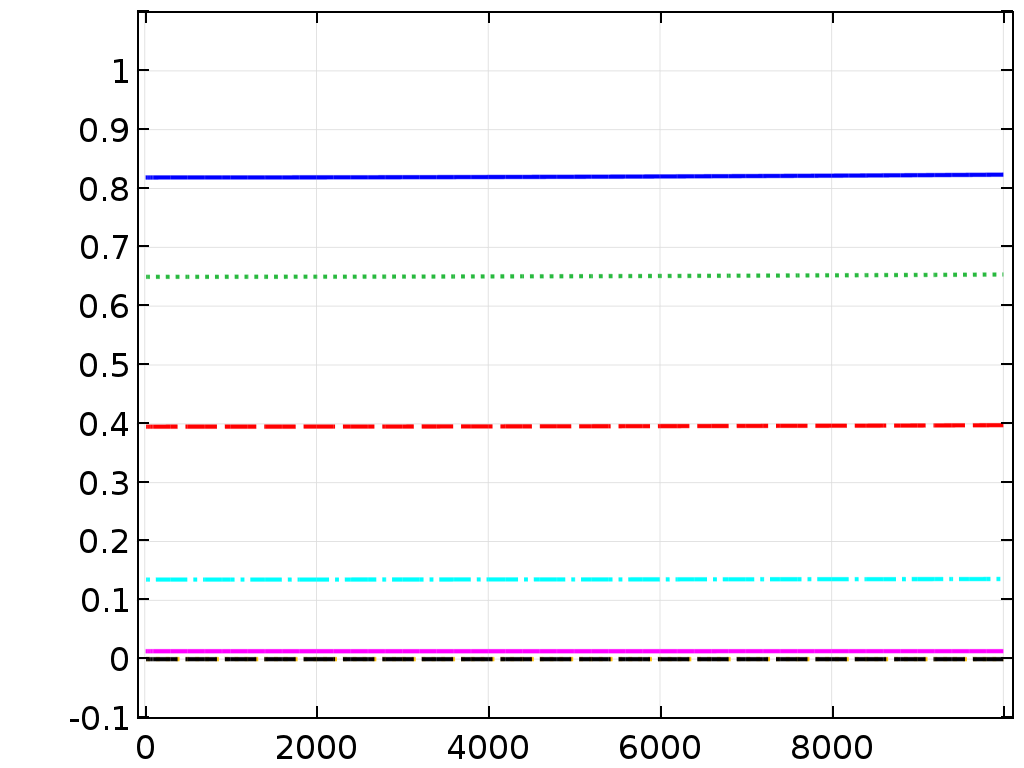}\includegraphics[width=.45\textwidth]{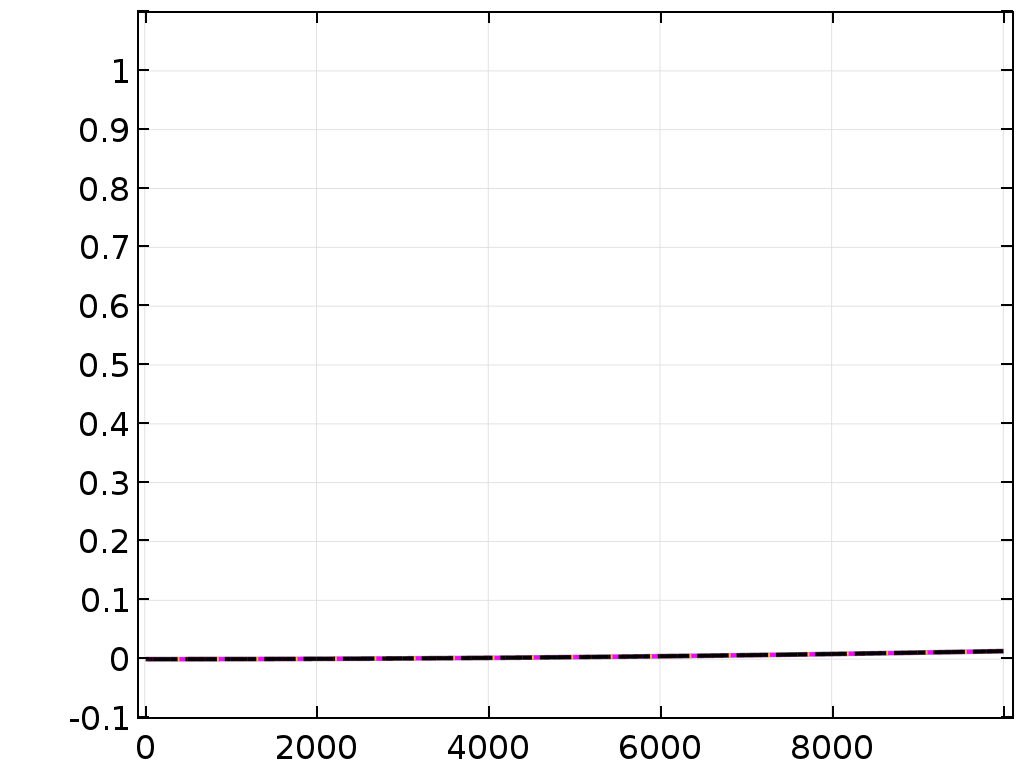}
	\caption{Scattering matrix coefficients magnitude (dimensionless) of the proposed acoustic isolator vs. frequency (in Hz), and for different control parameter values $d$. Top-left: $S_{11}$, top-right: $S_{12}$, bottom-left: $S_{21}$, and bottom-right: $S_{22}$.} 
	\label{fig:Smatrix}
	
\end{figure*}

The scattering matrix coefficients can be obtained, analytically or numerically. We illustrate each of these coefficient in the figure \ref{fig:Smatrix} by representing the magnitude (in dB) of each coefficient as a function of the frequency, for several values of $d$. It is clear that $S_{12}$ is very close to 1, and $S_{22}$ to 0, whatever the value of $d$ and the frequency, which illustrates the perfect transmission for backward waves. $S_{21}$ follows exactly the expected value exp($-L/d$), independently of the frequency. And the reflection coefficient for forward waves $S_{11}$ shows a frequency- and $d$-dependent trend: it decreases both with frequency and the value of the control parameter $d$.

\subsection{Isolation}
\begin{figure}
	\centering
	\includegraphics[width=8.6cm]{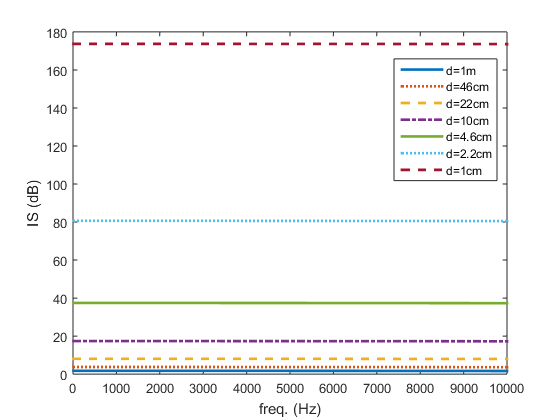}
	\caption{Isolation (in dB) vs frequency, for different values of the control parameter $d$.}
	\label{fig:IS}
\end{figure}

The isolation index, defined here as $IS = 20 log_{10}(|S_{12}/S_{21}|)$, characterizes the performance of the isolator. It is shown in figure \ref{fig:IS} as a function of frequency, for different values of control parameter $d$. For each simulation, with a fixed value of $d$, the isolation is constant over the whole frequency range. It follows perfectly the exponential attenuation factor $exp(L/d)$, resulting in an isolation index proportional to $L/d$. Hence, the required isolation for a given application can be specified and tuned accordingly.

\subsection{Stability}
\begin{figure}
	\centering
	\includegraphics[width=8.6cm]{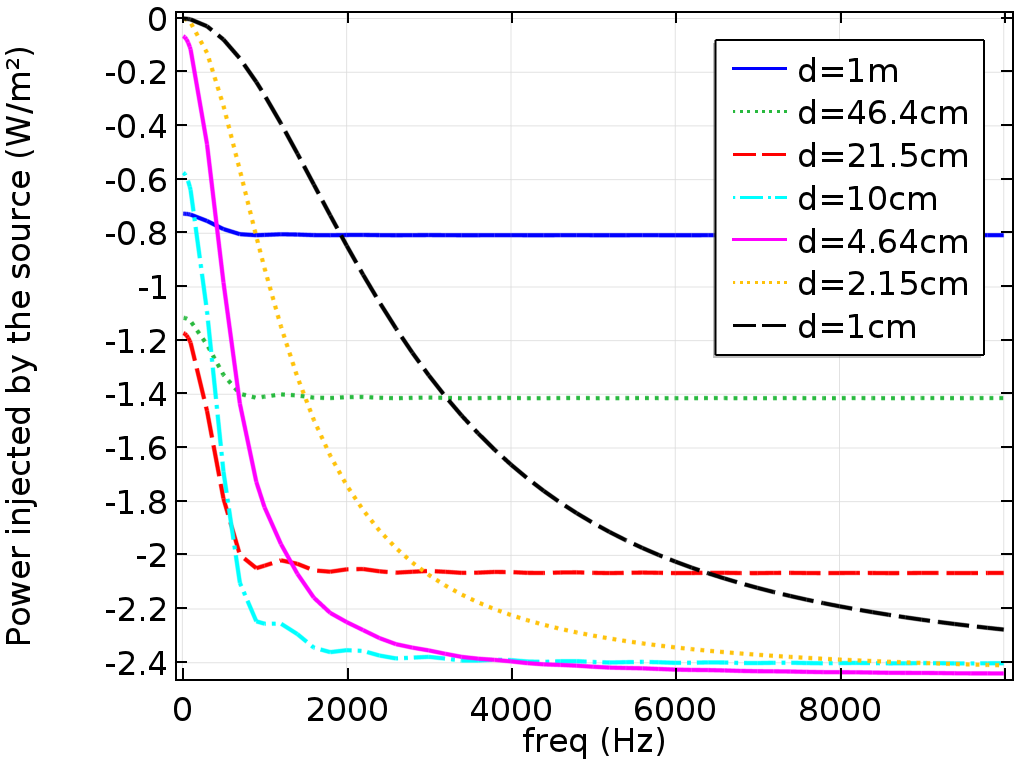}
	\caption{Total power injected by the source into the acoustic domain inside the isolator, as a function of the frequency, for different values of the control parameter $d$.}
	\label{fig:1D-Psource-freq}
\end{figure}

The power balance computed in post-processing shows that the distributed source never brings energy to the system, as expected. Moreover, the numerical results match perfectly the theoretical prediction obtained by integration of equation \eqref{eq:Psource} along the waveguide. Figure \ref{fig:1D-Psource-freq}, where the total power injected by the source $P_{source}(\w) = \int_{x=0}^{L}P(x,\w)dx$ is plotted against the frequency, for different values of the control parameter $d$.

\vskip\lineskip
As a summary, the proposed device provides a tunable isolator, with flat frequency response over at least 3 decades of bandwidth. Moreover, using $d$=$L$/10, an efficient acoustic isolator is obtained: the isolation is greater than 80dB, with a near-perfect transmission in the pass direction. The isolator presented here is as compact as $\lambda$/173 at 10Hz.

\section{Discussion}
The exceptional feature of the proposed device is its inherently broadband nature, and its performance at very low frequency, ie. in the deep sub-wavelength regime. Whereas most isolators or circulators are resonance-based and can exhibit very good isolation and low insertion losses on a very narrow band only, the proposed device is broadband by design.

\subsection{Discretization and transducer dynamics}
In this part we discuss the practical issues that affect the performance when building an actual acoustic isolator. In practice, a waveguide can be built with in-wall, flush mounted actuators and sensors. A proper control of each actuator, using sensors information, leads to the concept of programmable boundary condition as presented in the introduction. A first programmable boundary has been proposed under the form of an active acoustic metasurface\cite{karkar:2015a}, using the control approach proposed by \citet{rivet:2016}. Such a boundary can interact with the inner acoustic field and mimic the distributed source used in our 1D model, providing the condition corresponding to eq. \eqref{bndcontrol} is programmed onto this kind of active acoustic metamaterial.

Such discretization of the distributed source with unit-cells can lead to inaccuracies, which can in turn lead to loss of performance, or even loss of stability. Thus, it is important to discuss the effects of this discretization, as it might prevent the practical feasibility of such devices.

Two effects can impair the isolator performances: the use of a finite number of discrete sources, the finite length of the discrete sources.

While the first effect is easily compensable by adjusting the gain of the control, or equivalently adjusting parameter $d$, the second effect is less predictable, and probably more important in terms of performance loss. It mainly depends on the unit cell architecture. However, at 10kHz, the wavelength in free air is about 3.4cm, so it is anticipated that sources having a length of the order of 5mm should not have a significant impact on the performances up to that frequency. Conversely, sources of that size might not deliver enough acoustic flow at very low frequency. Hence a compromise on the bandwidth could come from the choice of transducers.

While the unit cells can be realized using miniature electro-dynamic loudspeakers, such devices are inherently band-pass systems, and possess their own low-cut and high-cut frequencies. However, Rivet et al. \cite{rivet:2016} showed that, using only one pressure sensor and a current source, very wide bandwidths can be obtained with an electro-dynamic loudspeaker: $Q\simeq 0.1$. In our case, choosing actuators with a resonance frequency around 1kHz would thus allow us to cover the whole frequency range of interest (10Hz-10kHz), without major alteration of the performances. Such actuators would have a typical size of a few cm, hence the experimental realization would actually suffer, in the high frequency range, of the limitation due to their size, and not their dynamics.


\section{Conclusion}
In this paper, we introduced the concept of programmable boundary condition, and we showed how an efficient, practical acoustic isolator can be designed and realized in the audible range, based on that concept. The proposed device was characterized by means of the scattering matrix, and numerical simulations confirm that such device realizes a one-way transmission of sound in ducts. Influence of the discretization and transducers' dynamics on the actual performance was assessed numerically. Future works will investigate the stability issue for discrete and finite-sized sources, and propose an experimental realization of this acoustic isolator.

\bibliographystyle{apsrev}
\bibliography{acoustic-diode}

\begin{thebibliography}{19}
\expandafter\ifx\csname natexlab\endcsname\relax\def\natexlab#1{#1}\fi
\expandafter\ifx\csname bibnamefont\endcsname\relax
  \def\bibnamefont#1{#1}\fi
\expandafter\ifx\csname bibfnamefont\endcsname\relax
  \def\bibfnamefont#1{#1}\fi
\expandafter\ifx\csname citenamefont\endcsname\relax
  \def\citenamefont#1{#1}\fi
\expandafter\ifx\csname url\endcsname\relax
  \def\url#1{\texttt{#1}}\fi
\expandafter\ifx\csname urlprefix\endcsname\relax\def\urlprefix{URL }\fi
\providecommand{\bibinfo}[2]{#2}
\providecommand{\eprint}[2][]{\url{#2}}

\bibitem[{\citenamefont{Aplet and Carson}(1964)}]{aplet:1964}
\bibinfo{author}{\bibfnamefont{L.}~\bibnamefont{Aplet}} \bibnamefont{and}
  \bibinfo{author}{\bibfnamefont{J.}~\bibnamefont{Carson}},
  \bibinfo{journal}{Applied Optics} \textbf{\bibinfo{volume}{3}},
  \bibinfo{pages}{544} (\bibinfo{year}{1964}).

\bibitem[{\citenamefont{Krause et~al.}(2008)\citenamefont{Krause, Renner, and
  Brinkmeyer}}]{krause:2008}
\bibinfo{author}{\bibfnamefont{M.}~\bibnamefont{Krause}},
  \bibinfo{author}{\bibfnamefont{H.}~\bibnamefont{Renner}}, \bibnamefont{and}
  \bibinfo{author}{\bibfnamefont{E.}~\bibnamefont{Brinkmeyer}},
  \bibinfo{journal}{Electronics Letters} \textbf{\bibinfo{volume}{44}},
  \bibinfo{pages}{691} (\bibinfo{year}{2008}).

\bibitem[{\citenamefont{Poulton et~al.}(2012)\citenamefont{Poulton, Pant,
  Byrnes, Fan, Steel, and Eggleton}}]{poulton:2012}
\bibinfo{author}{\bibfnamefont{C.~G.} \bibnamefont{Poulton}},
  \bibinfo{author}{\bibfnamefont{R.}~\bibnamefont{Pant}},
  \bibinfo{author}{\bibfnamefont{A.}~\bibnamefont{Byrnes}},
  \bibinfo{author}{\bibfnamefont{S.}~\bibnamefont{Fan}},
  \bibinfo{author}{\bibfnamefont{M.}~\bibnamefont{Steel}}, \bibnamefont{and}
  \bibinfo{author}{\bibfnamefont{B.~J.} \bibnamefont{Eggleton}},
  \bibinfo{journal}{Optics express} \textbf{\bibinfo{volume}{20}},
  \bibinfo{pages}{21235} (\bibinfo{year}{2012}).

\bibitem[{\citenamefont{Tocci et~al.}(1995)\citenamefont{Tocci, Bloemer,
  Scalora, Dowling, and Bowden}}]{tocci:1995}
\bibinfo{author}{\bibfnamefont{M.~D.} \bibnamefont{Tocci}},
  \bibinfo{author}{\bibfnamefont{M.~J.} \bibnamefont{Bloemer}},
  \bibinfo{author}{\bibfnamefont{M.}~\bibnamefont{Scalora}},
  \bibinfo{author}{\bibfnamefont{J.~P.} \bibnamefont{Dowling}},
  \bibnamefont{and} \bibinfo{author}{\bibfnamefont{C.~M.}
  \bibnamefont{Bowden}}, \bibinfo{journal}{Applied physics letters}
  \textbf{\bibinfo{volume}{66}}, \bibinfo{pages}{2324} (\bibinfo{year}{1995}).

\bibitem[{\citenamefont{Gallo et~al.}(2001)\citenamefont{Gallo, Assanto,
  Parameswaran, and Fejer}}]{gallo:2001}
\bibinfo{author}{\bibfnamefont{K.}~\bibnamefont{Gallo}},
  \bibinfo{author}{\bibfnamefont{G.}~\bibnamefont{Assanto}},
  \bibinfo{author}{\bibfnamefont{K.~R.} \bibnamefont{Parameswaran}},
  \bibnamefont{and} \bibinfo{author}{\bibfnamefont{M.~M.} \bibnamefont{Fejer}},
  \bibinfo{journal}{Applied Physics Letters} \textbf{\bibinfo{volume}{79}},
  \bibinfo{pages}{314} (\bibinfo{year}{2001}).

\bibitem[{\citenamefont{Lira et~al.}(2012)\citenamefont{Lira, Yu, Fan, and
  Lipson}}]{lira:2012}
\bibinfo{author}{\bibfnamefont{H.}~\bibnamefont{Lira}},
  \bibinfo{author}{\bibfnamefont{Z.}~\bibnamefont{Yu}},
  \bibinfo{author}{\bibfnamefont{S.}~\bibnamefont{Fan}}, \bibnamefont{and}
  \bibinfo{author}{\bibfnamefont{M.}~\bibnamefont{Lipson}},
  \bibinfo{journal}{Physical Review Letters} \textbf{\bibinfo{volume}{109}},
  \bibinfo{pages}{033901} (\bibinfo{year}{2012}).

\bibitem[{\citenamefont{Yu and Fan}(2009)}]{yu:2009}
\bibinfo{author}{\bibfnamefont{Z.}~\bibnamefont{Yu}} \bibnamefont{and}
  \bibinfo{author}{\bibfnamefont{S.}~\bibnamefont{Fan}},
  \bibinfo{journal}{Nature Photonics} \textbf{\bibinfo{volume}{3}},
  \bibinfo{pages}{91} (\bibinfo{year}{2009}).

\bibitem[{\citenamefont{Jalas et~al.}(2013)\citenamefont{Jalas, Petrov, Eich,
  Freude, Fan, Yu, Baets, Popovic, Melloni, Joannopoulos et~al.}}]{jalas:2013}
\bibinfo{author}{\bibfnamefont{D.}~\bibnamefont{Jalas}},
  \bibinfo{author}{\bibfnamefont{A.}~\bibnamefont{Petrov}},
  \bibinfo{author}{\bibfnamefont{M.}~\bibnamefont{Eich}},
  \bibinfo{author}{\bibfnamefont{W.}~\bibnamefont{Freude}},
  \bibinfo{author}{\bibfnamefont{S.}~\bibnamefont{Fan}},
  \bibinfo{author}{\bibfnamefont{Z.}~\bibnamefont{Yu}},
  \bibinfo{author}{\bibfnamefont{R.}~\bibnamefont{Baets}},
  \bibinfo{author}{\bibfnamefont{M.}~\bibnamefont{Popovic}},
  \bibinfo{author}{\bibfnamefont{A.}~\bibnamefont{Melloni}},
  \bibinfo{author}{\bibfnamefont{J.~D.} \bibnamefont{Joannopoulos}},
  \bibnamefont{et~al.}, \bibinfo{journal}{Nature Photonics}
  \textbf{\bibinfo{volume}{7}}, \bibinfo{pages}{579} (\bibinfo{year}{2013}).

\bibitem[{\citenamefont{Riedlinger}(1986)}]{brevet_diode}
\bibinfo{author}{\bibfnamefont{R.}~\bibnamefont{Riedlinger}},
  \emph{\bibinfo{title}{Acoustic diode}} (\bibinfo{year}{1986}),
  \bibinfo{note}{patent number: US4618796A}.

\bibitem[{\citenamefont{Liang et~al.}(2009)\citenamefont{Liang, Yuan, and
  Cheng}}]{liang:2009}
\bibinfo{author}{\bibfnamefont{B.}~\bibnamefont{Liang}},
  \bibinfo{author}{\bibfnamefont{B.}~\bibnamefont{Yuan}}, \bibnamefont{and}
  \bibinfo{author}{\bibfnamefont{J.-c.} \bibnamefont{Cheng}},
  \bibinfo{journal}{Physycal Review Letters} \textbf{\bibinfo{volume}{103}},
  \bibinfo{pages}{104301} (\bibinfo{year}{2009}).

\bibitem[{\citenamefont{Liang et~al.}(2010)\citenamefont{Liang, Guo, Tu, and
  Cheng}}]{liang:2010}
\bibinfo{author}{\bibfnamefont{B.}~\bibnamefont{Liang}},
  \bibinfo{author}{\bibfnamefont{X.~S.} \bibnamefont{Guo}},
  \bibinfo{author}{\bibfnamefont{J.}~\bibnamefont{Tu}}, \bibnamefont{and}
  \bibinfo{author}{\bibfnamefont{J.~C.} \bibnamefont{Cheng}},
  \bibinfo{journal}{Nature Materials} \textbf{\bibinfo{volume}{9}},
  \bibinfo{pages}{989} (\bibinfo{year}{2010}).

\bibitem[{\citenamefont{Boechler et~al.}(2011)\citenamefont{Boechler,
  Theocharis, and Daraio}}]{boechler:2011}
\bibinfo{author}{\bibfnamefont{N.}~\bibnamefont{Boechler}},
  \bibinfo{author}{\bibfnamefont{G.}~\bibnamefont{Theocharis}},
  \bibnamefont{and} \bibinfo{author}{\bibfnamefont{C.}~\bibnamefont{Daraio}},
  \bibinfo{journal}{Nature Materials} \textbf{\bibinfo{volume}{10}},
  \bibinfo{pages}{665} (\bibinfo{year}{2011}).

\bibitem[{\citenamefont{Popa and Cummer}(2014)}]{popa:2014}
\bibinfo{author}{\bibfnamefont{B.-I.} \bibnamefont{Popa}} \bibnamefont{and}
  \bibinfo{author}{\bibfnamefont{S.}~\bibnamefont{Cummer}},
  \bibinfo{journal}{Nature Communications} \textbf{\bibinfo{volume}{5}}
  (\bibinfo{year}{2014}).

\bibitem[{\citenamefont{Fleury et~al.}(2014)\citenamefont{Fleury, Sounas,
  Sieck, Haberman, and Al{\`u}}}]{Fleury:2014}
\bibinfo{author}{\bibfnamefont{R.}~\bibnamefont{Fleury}},
  \bibinfo{author}{\bibfnamefont{D.~L.} \bibnamefont{Sounas}},
  \bibinfo{author}{\bibfnamefont{C.~F.} \bibnamefont{Sieck}},
  \bibinfo{author}{\bibfnamefont{M.~R.} \bibnamefont{Haberman}},
  \bibnamefont{and} \bibinfo{author}{\bibfnamefont{A.}~\bibnamefont{Al{\`u}}},
  \bibinfo{journal}{Science} \textbf{\bibinfo{volume}{343}},
  \bibinfo{pages}{516} (\bibinfo{year}{2014}).

\bibitem[{\citenamefont{Fleury et~al.}(2015)\citenamefont{Fleury, Sounas, and
  Al\`u}}]{fleury:2015}
\bibinfo{author}{\bibfnamefont{R.}~\bibnamefont{Fleury}},
  \bibinfo{author}{\bibfnamefont{D.~L.} \bibnamefont{Sounas}},
  \bibnamefont{and} \bibinfo{author}{\bibfnamefont{A.}~\bibnamefont{Al\`u}},
  \bibinfo{journal}{Physical Review B} \textbf{\bibinfo{volume}{91}},
  \bibinfo{pages}{174306} (\bibinfo{year}{2015}).

\bibitem[{\citenamefont{Maznev et~al.}(2013)\citenamefont{Maznev, Every, and
  Wright}}]{Maznev:2013}
\bibinfo{author}{\bibfnamefont{A.}~\bibnamefont{Maznev}},
  \bibinfo{author}{\bibfnamefont{A.}~\bibnamefont{Every}}, \bibnamefont{and}
  \bibinfo{author}{\bibfnamefont{O.}~\bibnamefont{Wright}},
  \bibinfo{journal}{Wave Motion} \textbf{\bibinfo{volume}{50}},
  \bibinfo{pages}{776 } (\bibinfo{year}{2013}).

\bibitem[{\citenamefont{Karkar et~al.}(2015)\citenamefont{Karkar, Lissek,
  Collet, Ouisse, Matten, and Versaevel}}]{karkar:2015a}
\bibinfo{author}{\bibfnamefont{S.}~\bibnamefont{Karkar}},
  \bibinfo{author}{\bibfnamefont{H.}~\bibnamefont{Lissek}},
  \bibinfo{author}{\bibfnamefont{M.}~\bibnamefont{Collet}},
  \bibinfo{author}{\bibfnamefont{M.}~\bibnamefont{Ouisse}},
  \bibinfo{author}{\bibfnamefont{G.}~\bibnamefont{Matten}}, \bibnamefont{and}
  \bibinfo{author}{\bibfnamefont{M.}~\bibnamefont{Versaevel}}, in
  \emph{\bibinfo{booktitle}{Proceedings of INTER-NOISE 2015 -- 44th
  International Congress and Exposition on Noise Control Engineering}}
  (\bibinfo{year}{2015}).

\bibitem[{\citenamefont{Collet et~al.}(2009)\citenamefont{Collet, David, and
  Berthillier}}]{collet:2009}
\bibinfo{author}{\bibfnamefont{M.}~\bibnamefont{Collet}},
  \bibinfo{author}{\bibfnamefont{P.}~\bibnamefont{David}}, \bibnamefont{and}
  \bibinfo{author}{\bibfnamefont{M.}~\bibnamefont{Berthillier}},
  \bibinfo{journal}{The Journal of the Acoustical Society of America}
  \textbf{\bibinfo{volume}{125}}, \bibinfo{pages}{882} (\bibinfo{year}{2009}).

\bibitem[{\citenamefont{Rivet et~al.}(2017)\citenamefont{Rivet, Karkar, and
  Lissek}}]{rivet:2016}
\bibinfo{author}{\bibfnamefont{E.}~\bibnamefont{Rivet}},
  \bibinfo{author}{\bibfnamefont{S.}~\bibnamefont{Karkar}}, \bibnamefont{and}
  \bibinfo{author}{\bibfnamefont{H.}~\bibnamefont{Lissek}},
  \bibinfo{journal}{IEEE Transactions on Control Systems Technology}
  \textbf{\bibinfo{volume}{25}}, \bibinfo{pages}{63} (\bibinfo{year}{2017}).

\end{thebibliography}

\end{document}